\documentclass[prb,superscriptaddress,showpacs,floatfix,tightenlines,10pt,twocolumn]{revtex4}
\usepackage{amssymb}
\usepackage{amsmath}
\usepackage{graphicx}

\begin{document}

\title{Validity of the two-band model of bilayer and trilayer graphene in a
magnetic field}
\author{R. C\^{o}t\'{e}}
\affiliation{D\'{e}partement de physique, Universit\'{e} de Sherbrooke, Sherbrooke, Qu%
\'{e}bec, J1K 2R1, Canada}
\author{Manuel Barrette}
\affiliation{D\'{e}partement de physique, Universit\'{e} de Sherbrooke, Sherbrooke, Qu%
\'{e}bec, J1K 2R1, Canada}
\keywords{graphene bilayer, graphene trilayer, Landau levels,two-band model }
\pacs{73.21.-b,73.22.Dj,73.22.Pr}

\begin{abstract}
The eigenstates of an electron in the chiral two-dimensional electron gas
(C2DEG) formed in an AB-stacked bilayer or an ABC-stacked trilayer graphene
is a spinor with $4$ or $6$ components respectively. These components give
the amplitude of the wave function on the $4$ or $6$ carbon sites in the
unit cell of the lattice. In the tight-binding approximation, the
eigenenergies are thus found by diagonalizing a $4\times 4$ or a $6\times 6$
matrix. In the continuum approximation where the electron wave vector $%
k<<1/a_{0},$ with $a_{0}$ the lattice constant of the graphene sheets, a
common approximation is the two-band model\cite{McCann2006} (2BM)\ where the
eigenstates for the bilayer and trilayer systems are described by a
two-component spinor that gives the amplitude of the wave function on the
two sites with low-energy $\left\vert E\right\vert <<\gamma _{1}$ where $%
\gamma _{1}$ is the hopping energy between sites that are directly above one
another in adjacent layers. The 2BM has been used extensively to study the
phase diagram of the C2DEG in a magnetic field as well as its transport and
optical properties. In this paper, we use a numerical approach to compute
the eigenstates and Landau level energies of the full tight-binding model in
the continuum approximation and compare them with the prediction of the 2BM
when the magnetic field or an electrical bias between the outermost layers
is varied. Our numerical analysis shows that the 2BM is a good approximation
for bilayer graphene in a wide range of magnetic field and bias but mostly
for Landau level $M=0.$ The applicability of the 2BM in trilayer graphene,
even for level $M=0,$ is much more restricted. In this case, the 2BM\ fails
to reproduce some of the level crossings that occur between the sub-levels
of $M=0.$
\end{abstract}

\date{\today }
\maketitle

\section{INTRODUCTION}

Electrons in AB- or Bernal-stacked bilayer graphene (AB-BLG)\ and in
rhombohedral or ABC-stacked trilayer graphene (ABC-TLG) behave as a chiral
two-dimensional electron gas (C2DEG)\ of Dirac fermions\cite%
{McCann2006,BarlasRevue2012,Min2008} that has transport properties different
from those of the conventional 2DEG formed in semiconductor
heterostructures. For example, the Landau level spectrum in the simplest
(minimal) tight-binding model of a C2DEG is given by $E_{M}=$sgn$\left(
M\right) \frac{\alpha _{0}^{2}}{\gamma _{1}}\sqrt{\left\vert M\right\vert
\left( \left\vert M\right\vert +1\right) }$ for AB-BLG and $E_{M}=$sgn$%
\left( M\right) \frac{\alpha _{0}^{3}}{\gamma _{1}^{2}}\sqrt{\left(
\left\vert M\right\vert +2\right) \left( \left\vert M\right\vert +1\right)
\left\vert M\right\vert }$ for ABC-TLG where $M=0,\pm 1,\pm 2,...$ is the
Landau level index, $\gamma _{1}$ is the interlayer hopping amplitude, $%
\alpha _{0}=\sqrt{\frac{3}{2}}\frac{\gamma _{0}a_{0}}{\ell }$ (with $\gamma
_{0}$ the nearest-neighbor hopping amplitude in each graphene plane), $%
a_{0}=2.\,\allowbreak 46$ \AA\ is the lattice constant of graphene and $\ell
=\sqrt{\frac{\hslash c}{eB}}$ is the magnetic length. This spectrum leads to
the anomalous quantum Hall effect with conductivity $\sigma _{xy}=\pm g\frac{%
e^{2}}{h}(n+\frac{1}{2})$ where $n=0,1,2,...$ with $g=4$ for AB-BLG and $g=6$
for ABC-TLG. (For a review of BLG, see Refs. %
\onlinecite{CastroNetoRevue2009,AbergelRevue2010,GoerbigRevue2011,McCannRevue2013}%
).

In the above-mentionned graphene systems, each Landau level is fourfold
degenerate when counting spin and valley degrees of freedom. Level $M=0,$
however, is an exception. It has an extra \textit{orbital} degeneracy of two
for BLG and three for TLG. This, in turn, leads to an eightfold (BLG) or
twelvefold (TLG) degeneracy for that level. The extra degeneracy increases
the importance of Coulomb interaction in the C2DEG and new broken-symmetry
ground states (such as quantum Hall ferromagnetic states) can occur whose
experimental signature is a new plateau in the Hall conductivity.\cite%
{BarlasRevue2012} A very rich phase diagram for the C2DEG in these graphene
systems has been predicted at all integer filling factors in $M=0$ when the
magnetic field or the electrical potential difference (or bias $\Delta _{B}$%
) is applied between the two outermost layers.\cite%
{Lambert2013,Barlas2012,Rondeau2013}

A knowledge of the Landau level spectrum of the AB-BLG and ABC-TLG is
necessary to study the phase diagram and transport properties of the C2DEG.
The band structures of these two systems are well approximated by a
tight-binding Hamiltonian. In the minimal model, i.e. when keeping only the
two hopping parameters $\gamma _{0}$ and $\gamma _{1},$ the dispersion of
the four (six) $\pi $ bands in AB-BLG\ (ABC-TLG) is as shown in Fig. \ref%
{Fig1} for a wave vector $\mathbf{k}$ measured from either of the $K_{+}$ or 
$K_{-}$ valley point. Two low-energy ($\left\vert E\right\vert <<\gamma _{1}$%
) bands are separated from two or four high-energy bands by a gap $\Delta =$ 
$\gamma _{1}.$ An electronic state in each band is specified by a four-
(BLG) or six- (TLG) component spinor that gives the amplitude of the wave
function of each of the four or six sites in the unit cell of the lattice.

\begin{figure}[tbph]
\includegraphics[scale=1.0]{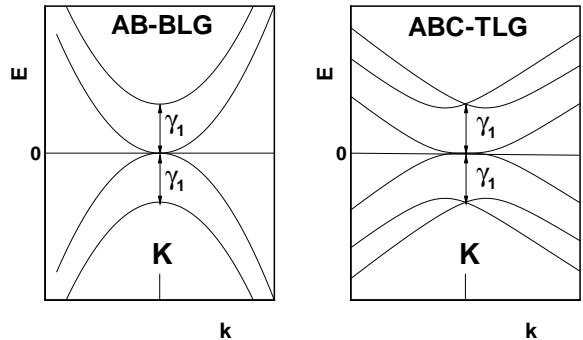}
\caption{Schematic drawing of the band structure of AB-stacked bilayer
graphene (left) and ABC-stacked trilayer graphene (right) in the minimal
model for wave vector $\mathbf{k}$ near the valley point $K.$ }
\label{Fig1}
\end{figure}

When the Fermi level is close to the degeneracy point of the two middle
bands in Fig. \ref{Fig1} and only the low-energy properties of the excited
states of the C2DEG are of interest, the complexity of these two graphene
systems can be substantially reduced by using an effective two-band model
(2BM) for BLG\cite{McCann2006,McCannRevue2013} and TLG.\cite%
{Koshino2009,Zhang2010} This model makes use of the fact that, for an
electronic state in one of the two low-energy bands, the amplitude of the
different components in the four- or six-component spinor is important only
on two low-energy sites. The high-energy sites can thus be integrated out.
The Hamiltonian is then reduced to a $2\times 2$ matrix and the eigenstates
are given by a two-component spinor. For completeness, the 2BM\ is reviewed
in Appendix A.

%%%%%%%%%%%%%%%%%%%%%%%%%%%%%%%%%%%%%%%%%%%%%%%%%%%%%%%%%%%%%%%%%%%%%%%%%
%%%%%%%%%%%%%%%%%%%%%%%%%%%%%%%%%%%%%%%%%%%%%%%%%%%%%%%%%%%%%%%%%%%%%%%%%%

The 2BM has been used extensively in the literature to study the phase
diagram of the C2DEG as well as its transport and optical properties.\cite%
{PhaseDiagram,Lambert2013,Rondeau2013,Barlas2012,Macdo2012} The phase
diagram, up to now, has been mostly studied as a function of magnetic field
and bias $\Delta _{B},$ and at integer filling of the eight or twelve
sub-Landau levels in $M=0$. To appreciate these studies or add corrections
such as screening where Landau levels $\left\vert M\right\vert >0$ are
important, or to study the phase diagram in higher Landau levels, one must
know the limits of validity of the 2BM. In the perturbation theory outlined
in Appendix A, the 2BM for BLG\ is estimated to be valid for low-energy
excitations $\left\vert E\right\vert <<\gamma _{1},$ and for magnetic field
such that $\gamma _{1}v_{3}/v_{0}^{2}<\hslash /\ell <\gamma _{1}/2v_{0}$
where $v_{i}=\frac{\sqrt{3}}{2}\frac{a_{0}}{\hslash }\left\vert \gamma
_{i}\right\vert $ and $\gamma _{3}$ is the so-called warping term. A similar
condition also applies in TLG.

In this work, we use a numerical approach to obtain a more precise
evaluation of the range of validity of the 2BM. A numerical computation of
the Landau level energies and eigenstates allows the inclusion of all
important hopping terms in the tight-binding model and thus give a more
realistic description of the levels that goes beyond that of the minimal
model\cite{Guinea2006,Yuan2011,Pereira2007} when values of the hopping
parameters typical of those found in the literature\cite%
{Castro2010,Zhang2010} are used. We compare the Landau level energies and
eigenstates in the full tight-binding model (in the continuum approximation)
with the predictions of the 2BM in a wide range of magnetic field and bias.
In our study, we pay particular attention to the eight (BLG) or twelve (TLG)
states in $M=0$ since it is for these levels that the 2BM is best suited.

This paper is organized in the following way. In Sec.\ II, we introduce the
AB-BLG and ABC-TLG systems as well as the tight-binding models that give the
Landau level spectrum of the full tight-binding model (4BM or 6BM) and of
the effective 2BM. In Sec. III, we discuss the effect of the different
hopping parameters $\gamma _{i}$ on the Landau level spectrum, we compare
the full and effective models for different values of the warping terms $%
\gamma _{2},\gamma _{3},$ bias and magnetic field and discuss the range of
validity of the effective model by comparing the eigenenergies and
eigenstates in the two models. We conclude in Sec. IV. In Appendix A, we
review the derivation of the 2BM.

\section{AB-BILAYER AND ABC-TRILAYER GRAPHENE}

\begin{figure}[tbph]
\includegraphics[scale=1.0]{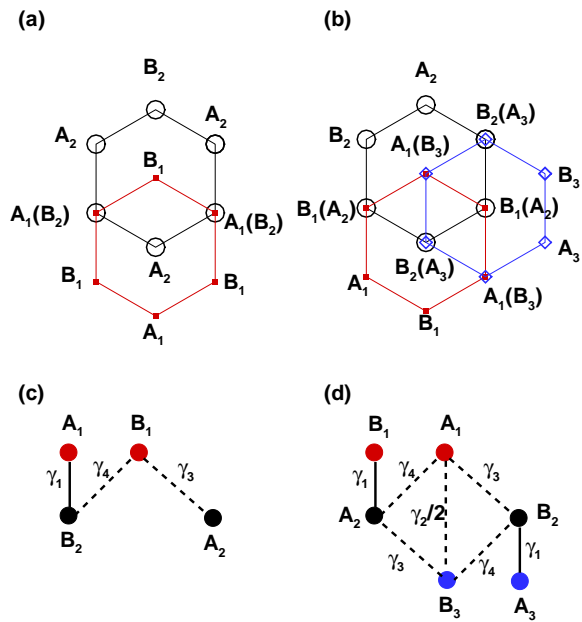}
\caption{(Color online) Crystal structure of the (a) Bernal-stacked bilayer
graphene and (b)\ ABC-stacked trilayer graphene. Definition of the hopping
parameters for the: (c) bilayer and (d) trilayer systems. }
\label{Fig2}
\end{figure}

The crystal structures of the AB-BLG and ABC-TLG (hereafter abbreviated as
BLG and TLG) are given in Fig. \ref{Fig2}(a),(b) and their tight-binding
parameters are defined in Fig. \ref{Fig2}(b),(c). (Note that a different
labelling is used for equivalent sites in the two structures.) The crystal
structure in each graphene layer is a honeycomb lattice that can be
described as a triangular Bravais lattice with a basis of two carbon atoms $%
A_{n}$ and $B_{n}$ where $n$ is the layer index. The triangular lattice
constant $a_{0}=2.\,\allowbreak 46$ \AA\ $=\sqrt{3}c,$\ where $c=1.42$ \AA\ %
is the separation between two adjacent carbon atoms. The unit cell of the
BLG (TLG)\ structure has four (six) carbon atoms. The distance between two
adjacent graphene layers is $d\approx 3.4$ \AA . The Brillouin zone of the
reciprocal lattice is hexagonal and has two nonequivalent valley points $%
\mathbf{K}_{\xi }=\left( \frac{2\pi }{a_{0}}\right) \left( \xi \frac{2}{3}%
,0\right) $ where $\xi =\pm $ is the valley index.

%%%%%%%%%%%%%%%%%%%%%%%%%%%%%%%%%%%%%%%%%%%%%%%%%%%%%%%%%%%%%%%%%%%%%%%%%
%%%%%%%%%%%%%%%%%%%%%%%%%%%%%%%%%%%%%%%%%%%%%%%%%%%%%%%%%%%%%%%%%%%%%%%%%%

The band structure of BLG or TLG is obtained with a tight-binding Hamiltonian%
\cite{McCann2006,Koshino2009,Zhang2010,Partoens2006}. The hopping parameters
common to both structures are:\ $\gamma _{0}$ the nearest-neighbor (NN)
hopping in each layer, $\gamma _{1}$ the interlayer hopping between carbon
atoms that are immediately above one another, $\gamma _{3}$ the interlayer
NN hopping term between carbon atoms of different sublattices, and $\gamma
_{4}$ the interlayer next NN hopping term between carbons atoms in the same
sublattice. In the trilayer, the extra hopping term $\gamma _{2}$ connects
the two low-energy sites $A_{1}$ and $B_{3}$ in Fig. \ref{Fig2}.

In the continuum approximation, the tight-binding Hamiltonian is expanded to
linear order in the wave vector $\mathbf{p}_{\pm }=\mathbf{k}-\mathbf{K}%
_{\pm }$. The effect of a magnetic field $\mathbf{B=}B\widehat{\mathbf{z}}$
perpendicular to the graphene layers is then taken into account by making
the Peierls substitution $\mathbf{p}\rightarrow \mathbf{P}=\mathbf{p}+e%
\mathbf{A}/\hslash c$ (with $e>0$ for an electron), where $\mathbf{A}$ is
the vector potential of the magnetic field.

\subsection{AB bilayer graphene}

In a magnetic field, the tight-binding Hamiltonian in the basis $\left\{
A_{1},B_{1},A_{2},B_{2}\right\} $ for valley $K_{-}$ and $\left\{
B_{2},A_{2},B_{1},A_{1}\right\} $ for valley $K_{+}$ is given by 
\begin{equation}
H_{\xi }=\left( 
\begin{array}{cccc}
\delta _{0}-\xi \frac{\Delta _{B}}{2} & \xi \alpha _{0}a & \xi \alpha
_{4}a^{\dag } & -\gamma _{1} \\ 
\xi \alpha _{0}a^{\dag } & -\xi \frac{\Delta _{B}}{2} & \xi \alpha _{3}a & 
\xi \alpha _{4}a^{\dag } \\ 
\xi \alpha _{4}a & \xi \alpha _{3}a^{\dag } & \xi \frac{\Delta _{B}}{2} & 
\xi \alpha _{0}a \\ 
-\gamma _{1} & \xi \alpha _{4}a & \xi \alpha _{0}a^{\dag } & \delta _{0}+\xi 
\frac{\Delta _{B}}{2}%
\end{array}%
\right) ,
\end{equation}%
where%
\begin{equation}
\alpha _{i}=\sqrt{\frac{3}{2}}\frac{a_{0}}{\ell }\gamma _{i},
\end{equation}%
and $\delta _{0}$ represents the difference in the crystal field between
sites $A_{1},B_{2}$ and $A_{2},B_{1}.$ The energy $\Delta _{B}$ is the
potential difference (or bias)\ between the two outermost layers due to an
external electric field perpendicular to the layers.

The Hamiltonian $H_{\xi }^{\left( 0\right) }$ obtained from $H_{\xi }$ by
setting $\alpha _{3}=0$ has eigenvalues $E_{N,j}^{\left( 0\right) }$ and
eigenspinors of the form%
\begin{equation}
\Psi _{N,j,X}^{\left( 0\right) }\left( \mathbf{r}\right) =\left( 
\begin{array}{c}
c_{N,j,1}h_{N-1,X}\left( \mathbf{r}\right) \\ 
c_{N,j,2}h_{N,X}\left( \mathbf{r}\right) \\ 
c_{N,j,3}h_{N-2,X}\left( \mathbf{r}\right) \\ 
c_{N,j,4}h_{N-1,X}\left( \mathbf{r}\right)%
\end{array}%
\right) ,  \label{psibilayer}
\end{equation}%
where $N\geq 0.$ In the Landau gauge $\mathbf{A}=\left( 0,Bx,0\right) $, the
functions $h_{n,X}\left( \mathbf{r}\right) $ are given by%
\begin{equation}
h_{n,X}\left( \mathbf{r}\right) =\frac{1}{\sqrt{L_{y}}}e^{-iXy/\ell
^{2}}\varphi _{n}\left( x-X\right) ,
\end{equation}%
where $L_{y}$ is the length of the 2DEG in the $y$ direction, $X$ the
guiding-center index, $\ell =\sqrt{\hslash c/eB}$ the magnetic length and $%
\varphi _{n}\left( x\right) $ are the wave functions of the one-dimensional
harmonic oscillator with quantum number $n=0,1,2,...$ The ladder operators
are defined such that $a\varphi _{n}\left( x\right) =-i\sqrt{n}\varphi
_{n-1}\left( x\right) $ and $a^{\dag }\varphi _{n}\left( x\right) =i\sqrt{n+1%
}\varphi _{n+1}\left( x\right) .$

The eigenvalues and coefficients $c_{N,j,k}$ for the eigenvectors are found
by diagonalizing the matrix%
\begin{equation}
F_{1}=\left( 
\begin{array}{cccc}
\delta _{0}-\xi \frac{\Delta _{B}}{2} & -\xi i\alpha _{0}g_{N} & \xi \alpha
_{4}ig_{N-1} & -\gamma _{1} \\ 
i\xi \alpha _{0}g_{N} & -\xi \frac{\Delta _{B}}{2} & 0 & \xi \alpha
_{4}ig_{N} \\ 
-i\xi \alpha _{4}g_{N-1} & 0 & \xi \frac{\Delta _{B}}{2} & -\xi i\alpha
_{0}g_{N-1} \\ 
-\gamma _{1} & -\xi i\alpha _{4}g_{N} & \xi i\alpha _{0}g_{N-1} & \delta
_{0}+\xi \frac{\Delta _{B}}{2}%
\end{array}%
\right) .
\end{equation}%
In $\Psi _{N,j,X}^{\left( 0\right) }\left( \mathbf{r}\right) $ and in the
above equation (and similar ones below), we \textit{define} $h_{N,X}\left( 
\mathbf{r}\right) =0$ if $N<0$ and 
\begin{equation}
g_{N}=\Theta \left( N\right) \sqrt{N},
\end{equation}%
where $\Theta \left( N\right) $ is the step function. $\,$For example, with $%
N=0,$ there can be only one eigenspinor ($j=1$) which is given by (we drop
hereafter the $\mathbf{r}$-dependency to simplify the notation) $\Psi
_{0,1,X}^{\left( 0\right) }=\left( 0,h_{0,X},0,0\right) $ and its energy is
precisely $E_{0,1}^{\left( 0\right) }=-\xi \frac{\Delta _{B}}{2}$ . For $%
N=1, $ there are three eigenstates ($j=1,2,3$) and four for $N>1$ (i.e. $%
j=1,2,...,4).$ The energies $E_{N,j}^{\left( 0\right) }$ and coefficients $%
c_{N,j,k}$ ($k=1,2,3,4)$ are independent of the guiding-center index $X$ so
that the Landau levels have the usual degeneracy $N_{\varphi }=S/2\pi \ell
^{2}$ where $S$ is the 2DEG area. (These energies and coefficients also
depend on the valley index $\xi $ but we do not indicate this in order to
simplify the notation.) In the basis $\left\{ \Psi _{N,j,X}^{\left( 0\right)
}\right\} $ of the eigenspinors of $H_{\xi }^{\left( 0\right) },$ the matrix
elements 
\begin{equation}
\left\langle \Psi _{N,j,X}^{\left( 0\right) }\left\vert H_{\xi }^{\left(
0\right) }\right\vert \Psi _{M,k,X^{\prime }}^{\left( 0\right)
}\right\rangle =\delta _{X,X^{\prime }}\delta _{N,M}\delta
_{j,k}E_{N,j}^{\left( 0\right) }.  \label{c_1}
\end{equation}

The effect of the warping term $\alpha _{3}$ is obtained by diagonalizing
the full matrix $H_{\xi }=H_{\xi }^{\left( 0\right) }+W_{\xi }$ where $%
W_{\xi }$ contains only the $\alpha _{3}$ terms and has the matrix elements%
\begin{eqnarray}
&&\left\langle \Psi _{N,j,X}^{\left( 0\right) }\left\vert W_{\xi
}\right\vert \Psi _{M,k,X^{\prime }}^{\left( 0\right) }\right\rangle
\label{c_2} \\
&=&-i\xi \alpha _{3}\sqrt{M-2}c_{N,j,2}^{\ast }c_{M,k,3}\delta
_{N,M-3}\delta _{X,X^{\prime }}  \notag \\
&&+i\xi \alpha _{3}\sqrt{M+1}c_{N,j,3}^{\ast }c_{M,k,2}\delta _{N,M+3}\delta
_{X,X^{\prime }}.  \notag
\end{eqnarray}%
This perturbation couples the eigenspinors $\Psi _{N,j,X}^{\left( 0\right) }$
with the same $X$ and, in view of Eq. (\ref{c_1}), the matrix to diagonalize
is the same for all $X^{\prime }s.$ We thus drop the $X$ index hereafter. If
the highest Landau level kept in the calculation is $N_{\max },$ then the
matrix for $H_{\xi }$ to diagonalize has order $4N_{\max }.$

The two-band model for bilayer graphene\cite{McCann2006,McCannRevue2013} can
be obtained in a number of ways. We describe one such way in Appendix A. In
the basis $\left( B_{1},A_{2}\right) $\ for $K_{-}$\ and $\left(
A_{2},B_{1}\right) $\textbf{\ }for $K_{+},$ the effective two-band
Hamiltonian is%
\begin{equation}
\widetilde{H}_{\mathbf{\xi }}=\left( 
\begin{array}{cc}
-\xi \frac{\Delta _{B}}{2}+\zeta _{+}a^{\dag }a & \frac{\alpha _{0}^{2}}{%
\gamma _{1}}\left( a^{\dag }\right) ^{2} \\ 
\frac{\alpha _{0}^{2}}{\gamma _{1}}a^{2} & \xi \frac{\Delta _{B}}{2}+\zeta
_{-}aa^{\dag }%
\end{array}%
\right) ,  \label{h2bm}
\end{equation}%
where%
\begin{eqnarray}
\beta &=&\frac{\alpha _{0}}{\gamma _{1}}, \\
\zeta &=&2\beta \alpha _{4}+\beta ^{2}\delta _{0}, \\
\zeta _{\pm } &=&\zeta \pm \xi \beta ^{2}\Delta _{B},
\end{eqnarray}%
The 2BM, as we define it here, does not include the warping term $\gamma
_{3}.$

The eigenspinors of the 2BM are of the form%
\begin{equation}
\widetilde{\Psi }_{N,j}=\left( 
\begin{array}{c}
c_{N,j,1}h_{N} \\ 
c_{N,j,2}h_{N-2}%
\end{array}%
\right) .
\end{equation}%
The $N=0$ eigenvector (with $j=1$) is $\widetilde{\Psi }_{0,1}=\left(
h_{0},0\right) $ and has energy%
\begin{equation}
\widetilde{E}_{0,1}^{\left( 0\right) }=-\xi \frac{\Delta _{B}}{2}.
\label{uno}
\end{equation}%
For $N=1,$ the eigenvector (with $j=1$) is $\widetilde{\Psi }_{1,1}=\left(
h_{1},0\right) $ with energy 
\begin{equation}
\widetilde{E}_{1,1}^{\left( 0\right) }=-\xi \frac{\Delta _{B}}{2}+\zeta _{+}.
\label{duo}
\end{equation}%
The Landau levels are obtained by diagonalizing the matrix 
\begin{equation}
F_{2}=\left( 
\begin{array}{cc}
-\xi \frac{\Delta _{B}}{2}+\zeta _{+}g_{N}^{2} & -\frac{\alpha _{0}^{2}}{%
\gamma _{1}}g_{N}g_{N-1} \\ 
-\frac{\alpha _{0}^{2}}{\gamma _{1}}g_{N}g_{N-1} & \xi \frac{\Delta _{B}}{2}%
+\zeta _{-}g_{N-1}^{2}%
\end{array}%
\right) .  \label{F2}
\end{equation}%
For $N>1$, the index $j=1,2.$

In the minimal model ($\gamma _{0},\gamma _{1}\neq 0$ only), the Landau
level spectrum is given by $E_{M}=$sgn$\left( M\right) \frac{\alpha _{0}^{2}%
}{\gamma _{1}}\sqrt{\left\vert M\right\vert \left( \left\vert M\right\vert
+1\right) }.$ The $N=0$ and $N=1$ states are degenerate and are part of the
Landau level $M=0.$ Below we refer to these states as the states $n=0$ and $%
n=1$ of $M=0$ where $n$ is the \textit{orbital} index in the eigenspinor $%
\left( h_{n},0\right) .$ The index $N\geq 0$ that we use to classify the
Landau levels in our numerical analysis is different from the Landau level $%
M.$ For $N\geq 2$ in Eq. (\ref{F2}), the relation is $M=\pm \left(
N-1\right) .$

\subsection{ABC-stacked trilayer graphene}

We repeat the above procedure to get the Landau levels of the ABC-TLG. The
tight-binding Hamiltonian in the basis $\left(
A_{1},B_{3},B_{1},A_{2},B_{2},A_{3}\right) $ for $K_{+}$ and $\left(
B_{3},A_{1},A_{3},B_{2},A_{2},B_{1}\right) $ for $K_{-}$ is%
\begin{equation}
H_{\xi }=\left( 
\begin{array}{cccccc}
\xi \frac{\Delta _{B}}{2}+\delta _{0} & \frac{\gamma _{2}}{2} & \xi \alpha
_{0}a & \xi \alpha _{4}a & \xi \alpha _{3}a^{\dag } & 0 \\ 
\frac{\gamma _{2}}{2} & -\xi \frac{\Delta _{B}}{2}+\delta _{0} & 0 & \xi
\alpha _{3}a & \xi \alpha _{4}a^{\dag } & \xi \alpha _{0}a^{\dag } \\ 
\xi \alpha _{0}a^{\dag } & 0 & \xi \frac{\Delta _{B}}{2} & \gamma _{1} & \xi
\alpha _{4}a & 0 \\ 
\xi \alpha _{4}a^{\dag } & \xi \alpha _{3}a^{\dag } & \gamma _{1} & 0 & \xi
\alpha _{0}a & \xi \alpha _{4}a \\ 
\xi \alpha _{3}a & \xi \alpha _{4}a & \xi \alpha _{4}a^{\dag } & \xi \alpha
_{0}a^{\dag } & 0 & \gamma _{1} \\ 
0 & \xi \alpha _{0}a & 0 & \xi \alpha _{4}a^{\dag } & \gamma _{1} & -\xi 
\frac{\Delta _{B}}{2}%
\end{array}%
\right) ,  \label{tri}
\end{equation}%
where the energy $\delta _{0}$ represents the difference in the crystal
field between sites $A_{1},B_{3}$ and sites $A_{2},A_{3},B_{1},B_{2}.$ (In
Eq. (\ref{tri}), it is assumed that the middle layer is at zero potential
and that $\Delta _{B}$ is the potential difference between the two outermost
layers.)

We define $H_{\xi }^{\left( 0\right) }$ as the Hamiltonian with $\gamma _{3}$
and $\gamma _{2}$ set to zero. Its eigenspinors are of the form 
\begin{equation}
\Psi _{N,j}^{\left( 0\right) }=\left( 
\begin{array}{c}
c_{N,j,1}h_{N-3} \\ 
c_{N,j,2}h_{N} \\ 
c_{N,j,3}h_{N-2} \\ 
c_{N,j,4}h_{N-2} \\ 
c_{N,j,5}h_{N-1} \\ 
c_{N,j,6}h_{N-1}%
\end{array}%
\right) .  \label{psitrilayer}
\end{equation}%
There is only one solution for $N=0$ (with the energy $E_{0,1}^{\left(
0\right) }=-\xi \frac{\Delta _{B}}{2}+\delta _{0}$ and $j=1$), three
solutions for $N=1$ $(j=1,2,3),$ five solutions for $N=2$ (i.e. $j=1,2,...,5$%
) and six for $N>2$ (i.e. $j=1,2,...,6$). The matrix to diagonalize is

\begin{widetext}

\begin{equation}
F_{3}=\left( 
\begin{array}{cccccc}
\xi \frac{\Delta _{B}}{2}+\delta _{0} & 0 & -i\xi \alpha _{0}g_{N-2} & -i\xi
\alpha _{4}g_{N-2} & 0 & 0 \\ 
0 & -\xi \frac{\Delta _{B}}{2}+\delta _{0} & 0 & 0 & i\xi \alpha _{4}g_{N} & 
i\xi \alpha _{0}g_{N} \\ 
i\xi \alpha _{0}g_{N-2} & 0 & \xi \frac{\Delta _{B}}{2} & \gamma _{1} & 
-i\xi \alpha _{4}g_{N-1} & 0 \\ 
i\xi \alpha _{4}g_{N-2} & 0 & \gamma _{1} & 0 & -i\xi \alpha _{0}g_{N-1} & 
-i\xi \alpha _{4}g_{N-1} \\ 
0 & -i\xi \alpha _{4}g_{N} & i\xi \alpha _{4}g_{N-1} & i\xi \alpha
_{0}g_{N-1} & 0 & \gamma _{1} \\ 
0 & -i\xi \alpha _{0}g_{N} & 0 & i\xi \alpha _{4}g_{N-1} & \gamma _{1} & 
-\xi \frac{\Delta _{B}}{2}%
\end{array}%
\right) .
\end{equation}

\end{widetext}

The Landau levels are obtained by diagonalizing the full matrix $H_{\mathbf{%
\xi }}=H_{\mathbf{\xi }}^{\left( 0\right) }+W_{\mathbf{\xi }}$ where $W_{%
\mathbf{\xi }}$ contains only the elements of $H_{\mathbf{\xi }}$ with $%
\gamma _{2}$ and $\alpha _{3}.$ The matrix elements of $W_{\mathbf{\xi }}$
in the basis $\left\{ \Psi _{N,j}^{\left( 0\right) }\right\} $ of the
eigenvectors of $H_{\xi }^{\left( 0\right) }$ are given by%
\begin{eqnarray}
&&\left\langle \Psi _{N,j}^{\left( 0\right) }\left\vert W_{\xi }\right\vert
\Psi _{M,k}^{\left( 0\right) }\right\rangle \\
&=&\frac{\gamma _{2}}{2}\left( \delta _{N-3,M}c_{N,j,1}^{\ast
}c_{M,k,2}+\delta _{N+3,M}c_{N,j,2}^{\ast }c_{M,k,1}\right)  \notag \\
&&-\delta _{N+3,M}i\xi \alpha _{3}\left( c_{N,j,2}^{\ast
}c_{M,k,4}g_{M-2}+c_{N,j,5}^{\ast }c_{M,k,1}g_{M-3}\right)  \notag \\
&&+\delta _{N-3,M}i\xi \alpha _{3}\left( c_{N,j,4}^{\ast
}c_{M,k,2}g_{M+1}+c_{N,j,1}^{\ast }c_{M,k,5}g_{M}\right) .  \notag
\end{eqnarray}%
If the highest Landau level kept in the calculation is $N_{\max },$ the
matrix to diagonalize has order $6\left( N_{\max }-2\right) +9.$

For the two-band model, the basis is $\left( A_{1},B_{3}\right) $ for $%
H_{K_{+}}$ and $\left( B_{3},A_{1}\right) $ for $H_{K_{-}}$ and the
Hamiltonian is given by 
\begin{equation}
\widetilde{H}_{\xi }=\left( 
\begin{array}{cc}
\delta _{0}+\xi \frac{\Delta _{B}}{2}-\widetilde{\zeta }_{+}aa^{\dag } & \xi 
\frac{\alpha _{0}^{3}}{\gamma _{1}^{2}}a^{3} \\ 
\xi \frac{\alpha _{0}^{3}}{\gamma _{1}^{2}}\left( a^{\dag }\right) ^{3} & 
\delta _{0}-\xi \frac{\Delta _{B}}{2}-\widetilde{\zeta }_{-}a^{\dag }a%
\end{array}%
\right) .
\end{equation}%
It does not include the $\gamma _{2}$ and $\gamma _{3}$ couplings. In this
equation, 
\begin{equation}
\widetilde{\zeta }_{\pm }=\zeta \pm \frac{1}{2}\xi \beta ^{2}\Delta _{B}.
\end{equation}

The eigenvectors of $\widetilde{H}_{\xi }$ have the form%
\begin{equation}
\widetilde{\Psi }_{N,j}=\left( 
\begin{array}{c}
c_{N,j,1}h_{N-3} \\ 
c_{N,j,2}h_{N}%
\end{array}%
\right) .
\end{equation}%
There is one solution only (i.e. $j=1$) for $N=0,1,2$ which are $\widetilde{%
\Psi }_{0,1}=\left( 0,h_{0}\right) ,\widetilde{\Psi }_{1,1}=\left(
0,h_{1}\right) ,\widetilde{\Psi }_{2,1}=\left( 0,h_{2}\right) $ with energy%
\begin{equation}
\widetilde{E}_{0,1}=\delta _{0}-\xi \frac{\Delta _{B}}{2},  \label{e1}
\end{equation}%
\begin{eqnarray}
\widetilde{E}_{1,1} &=&\left( 1-\beta ^{2}\right) \delta _{0}  \label{e2} \\
&&-\xi \left( 1-\beta ^{2}\right) \frac{\Delta _{B}}{2}-2\beta \alpha _{4}, 
\notag
\end{eqnarray}%
and%
\begin{eqnarray}
\widetilde{E}_{2,1} &=&\left( 1-2\beta ^{2}\right) \delta _{0}  \label{e3} \\
&&-\xi \left( 1-2\beta ^{2}\right) \frac{\Delta _{B}}{2}-4\beta \alpha _{4}.
\notag
\end{eqnarray}%
For $N>2,$ there are two solutions and the index $j=1,2.$

As in BLG, these three solutions are degenerate in the minimal model and
they belong to Landau level $M=0.$ Below, we refer to them as the orbital
states $n=0,1,2$ of $M=0.$ The Landau level quantization in the minimal
model is $E_{M}=$ sgn$\left( M\right) \frac{\alpha _{0}^{3}}{\gamma _{1}^{2}}%
\sqrt{\left( \left\vert M\right\vert +2\right) \left( \left\vert
M\right\vert +1\right) \left\vert M\right\vert }$ while, for $\widetilde{H}%
_{\xi },$ the Landau level spectrum is obtained by diagonalizing the matrix 
\begin{equation}
F_{4}=\left( 
\begin{array}{cc}
\delta _{0}+\xi \frac{\Delta _{B}}{2}-\widetilde{\zeta }_{+}g_{N-2}^{2} & 
i\xi \frac{\alpha _{0}^{3}}{\gamma _{1}^{2}}g_{N}g_{N-1}g_{N-2} \\ 
-i\xi \frac{\alpha _{0}^{3}}{\gamma _{1}^{2}}g_{N}g_{N-1}g_{N-2} & \delta
_{0}-\xi \frac{\Delta _{B}}{2}-\widetilde{\zeta }_{-}g_{N}^{2}%
\end{array}%
\right) .
\end{equation}%
With $N\geq 3$ in $F_{4},$ we obtain the energy of the Landau levels $M=\pm
\left( N-2\right) .$

\section{NUMERICAL RESULTS}

Each of the four or six bands in the band structure shown in Fig. \ref{Fig1}
leads to a set of Landau levels. Since we are interested in comparing the
full four-band (4BM) or six-band (6BM)\ models with the 2BM, we need to
consider only the Landau levels originating from the two low-energy bands.
In addition, the Landau level spectrum has the symmetry $E_{N,j}\left( \xi
,\Delta _{B}\right) =E_{N,j}\left( -\xi ,-\Delta _{B}\right) $ and so we can
further restrict our analysis to the spectrum in one valley. We choose the
valley $K_{+}$. For the numerical calculations, we use the parameters\textbf{%
: }$\gamma _{0}=3.16$ eV, $\gamma _{1}=0.502$ eV, $\gamma _{2}=-0.0171$ eV, $%
\gamma _{3}=-0.377$ eV, $\gamma _{4}=-0.099$ eV, $\delta _{0}=-0.0014$ eV
for TLG\cite{Zhang2010} and $\gamma _{0}=3.12$ eV, $\gamma _{1}=0.39$ eV, $%
\gamma _{3}=0.29$ eV, $\gamma _{4}=0.12$ eV, $\delta _{0}=0.0156$ eV for BLG%
\cite{Castro2010,Nilsson2008}. These parameters are not all precisely known
and can be affected by correlation effects and substrate.\cite{Gruneis2008}
However, small changes from these values should not affect significantly our
conclusions concerning the validity of the 2BM.

\subsection{Analytical criteria for the validity of the two-band model}

For the 2BM to be a valid approximation, the energy $\left\vert E\right\vert 
$ of the Landau levels must be much smaller than the inter-layer hopping
energy $\gamma _{1}$ i.e. smaller than the energy of the higher-energy
bands. In the minimal model $E\sim \alpha _{0}^{2}/\gamma _{1}$ for BLG and $%
E\sim \alpha _{0}^{3}/\gamma _{1}^{2}$ for TLG. If we require that $E<\gamma
_{1}/4$, then the condition of validity of the 2BM\cite{McCann2006} is $%
\hslash /\ell <\gamma _{1}/2v_{0}$\ (BLG) and $\hslash /\ell <\gamma
_{1}/2^{4/3}v_{0}$\ (TLG), where $v_{i}=\frac{\sqrt{3}}{2}\frac{a_{0}}{%
\hslash }\left\vert \gamma _{i}\right\vert .$ For our choice of parameters,
this condition implies that $B\lesssim 56$ T for BLG and TLG.

The warping terms $\gamma _{2},\gamma _{3}$ can also be included in the 2BM.
For BLG, this means adding to the Hamiltonian in Eq. (\ref{h2bm}) the term 
\begin{equation}
\widetilde{H}_{w,\mathbf{\xi }}=\xi \alpha _{3}\left( 
\begin{array}{cc}
0 & a \\ 
a^{\dag } & 0%
\end{array}%
\right)  \label{h2bmwarp}
\end{equation}%
and computing the eigenvalues numerically. For TLG, the warping term $\gamma
_{3}$ contributes\cite{Koshino2009} a term of order $2\alpha _{0}\alpha
_{3}/\gamma _{1}.$ If we require the warping term energy to be smaller than
the typical Landau level energy, then we need $\hslash /\ell >\gamma
_{1}v_{3}/v_{0}^{2}$ for BLG and $\hslash /\ell >2\gamma _{1}v_{3}/v_{0}^{2}$
for TLG. With our choice of parameters, this implies $B\gtrsim 2$ T for BLG
and $B\gtrsim $ $21$ T for TLG. In TLG, we arrive at approximately the same
magnetic field if we require that $\left\vert \gamma _{2}\right\vert
/2<\alpha _{0}^{3}/\gamma _{1}^{2},$ where $\left\vert \gamma
_{2}\right\vert /2$ is the typical energy contribution for the hopping term $%
\gamma _{2}.$

\subsection{Effect of the hopping terms $\protect\gamma _{2},\protect\gamma %
_{3}$}

We start our analysis by looking at the effect of the hopping terms $\gamma
_{2}$ and $\gamma _{3}$ that couple the eigenspinors $\Psi _{N,j}^{\left(
0\right) }$ together. For an AB-BLG, the effect of the warping term $\gamma
_{3}$ is discussed in Ref. \onlinecite{McCann2006} but within the 2BM. Fig.
3 of that paper shows how the $\gamma _{3}$ term couples the Landau levels
in the 2BM when $\gamma _{4},\delta _{0}=0$. For small $\alpha =\gamma
_{3}/\gamma _{0},$ the Landau level spectrum is almost independent of $%
\gamma _{3}$ while, at larger value of $\alpha ,$ groups of four consecutive
levels become degenerate, each group being separated from the next by two
nondegenerate Landau levels. In BLG, the value of $\gamma _{3}/\gamma
_{0}\approx 0.09$ is such that $\gamma _{3}$ has very little effect on the
spectrum if the magnetic field is large enough i.e. for $B\gtrsim 1$ T. This
is in good agreement with the analytical result given previously.

\begin{figure}[tbph]
\includegraphics[scale=1.0]{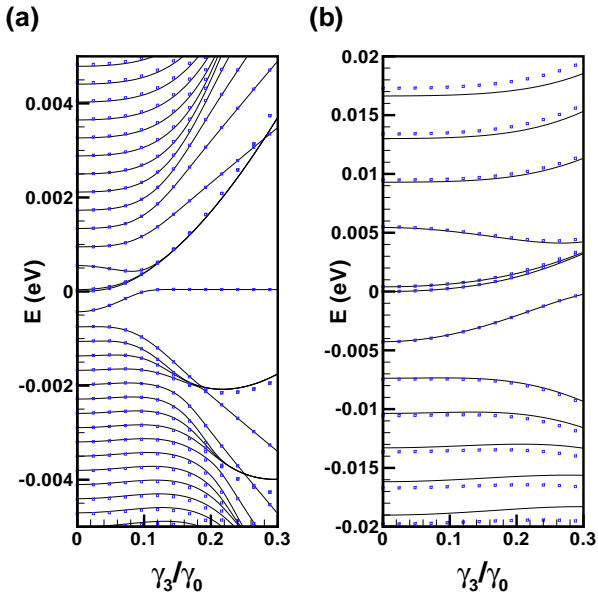}
\caption{Landau level spectrum as a function of the ratio $\protect\gamma %
_{3}/\protect\gamma _{0}$ at zero bias in the four-band (lines) and two-band
(blue squares) model of BLG\ for (a) $B=0.1$ T and (b) $B=1.0$ T. The
two-band model considered in this figure includes the warping term $\protect%
\gamma _{3}.$}
\label{Fig3}
\end{figure}

Figure \ref{Fig3} of our paper shows our numerical results for the effect of 
$\gamma _{3}$ on the Landau level spectrum but calculated in the 4BM. We
reach the same conclusion as in Ref. \onlinecite{McCann2006} i.e. the
warping term affects the spectrum only at very small magnetic field. For $%
B\gtrsim 1$ T, the effect of $\gamma _{3}$ is already very small. The
dispersion of the levels with $\gamma _{3}/\gamma _{0},$ however, is
qualitatively different than that of the 2BM. In the 4BM, groups of three
instead of four consecutive levels become degenerate. We can reproduce this
behavior with the 2BM\ if we include $\gamma _{3}$ \textit{and} $\gamma _{4}$
in the two-band Hamiltonian i.e. if we add to Eq. (\ref{h2bm}) the warping
term of Eq. (\ref{h2bmwarp}) and solve using the procedure described in Sec.
II. The dispersion of the resulting levels is shown by the blue squares in
Fig. \ref{Fig3}. Note that, in this figure, the magnetic field is small and
so the 4BM and 2BM (with warping)\ are in good agreement.

%%%%%%%%%%%%%%%%%%%%%%%%%%%%%%%%%%%%%%%%%%%%%%%%%%%%%%%%%%%%%%%%%%%%%%%%%
%%%%%%%%%%%%%%%%%%%%%%%%%%%%%%%%%%%%%%%%%%%%%%%%%%%%%%%%%%%%%%%%%%%%%%%%%%

\begin{figure}[tbph]
\includegraphics[scale=1.0]{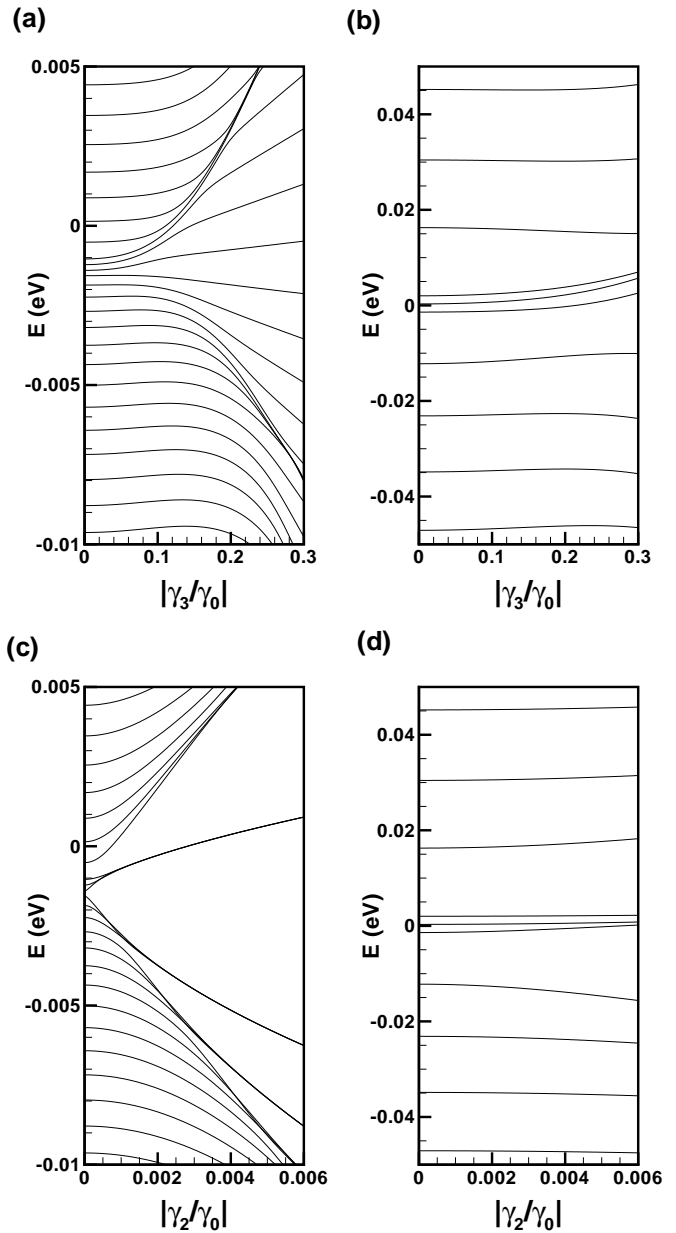}
\caption{Landau level spectrum in TLG\ at zero bias in the six-band model as
a function of the ratio $\left\vert \protect\gamma _{3}/\protect\gamma %
_{0}\right\vert $ with $\protect\gamma _{2}=0$ for (a) $B=1$ T and (b) $B=10$
T and as a function of the ratio $\left\vert \protect\gamma _{2}/\protect%
\gamma _{0}\right\vert $ with $\protect\gamma _{3}=0$ for (c) $B=1$ T and
(d) $B=10$ T.}
\label{Fig4}
\end{figure}

In the 6BM of TLG, the $\gamma _{2}$ term leads to Landau levels that are
degenerate by groups of three at small magnetic field (see also Fig. \ref%
{Fig9} (d) below). We thus set this term to zero in order to study the
effect of the warping term $\gamma _{3}$. At field $B=0.1$ T, Fig. \ref{Fig4}%
\ (a) shows that the warping term leads to an important modification of the
Landau level spectrum for $\left\vert \gamma _{3}/\gamma _{0}\right\vert
=0.12$ corresponding to the estimated value of the ratio of $\gamma _{3}$
and $\gamma _{0}$ in TLG. At the larger magnetic field $B=10$ T, Fig. \ref%
{Fig4}\ (b) shows that the effect of $\gamma _{3}$ is still noticeable. It
becomes negligible at a magnetic field $B\approx 20$ T (this can also be
seen in Fig. \ref{Fig5} (d) below).

%%%%%%%%%%%%%%%%%%%%%%%%%%%%%%%%%%%%%%%%%%%%%%%%%%%%%%%%%%%%%%%%%%%%%%%%%
%%%%%%%%%%%%%%%%%%%%%%%%%%%%%%%%%%%%%%%%%%%%%%%%%%%%%%%%%%%%%%%%%%%%%%%%%%

Figure \ref{Fig4}\ (c),(d) shows the same analysis but with $\gamma _{3}=0$
and varying $\left\vert \gamma _{2}/\gamma _{0}\right\vert .$ Again, the
spectrum is strongly influenced by $\gamma _{2}$ at low field. It becomes
independent of $\gamma _{2}$ at a magnetic field of the order of $B=20$ T
for our choice of hopping parameters i.e. $\left\vert \gamma _{2}/\gamma
_{0}\right\vert =0.0054.$ This value is again in agreement with the
analytical result given above. For TLG, it is necessary to go to much larger
magnetic fields than in BLG to avoid the Landau level degeneracy caused by $%
\gamma _{2}$ and $\gamma _{3}.$

\subsection{Dispersion with magnetic field}

\begin{figure}[tbph]
\includegraphics[scale=1.0]{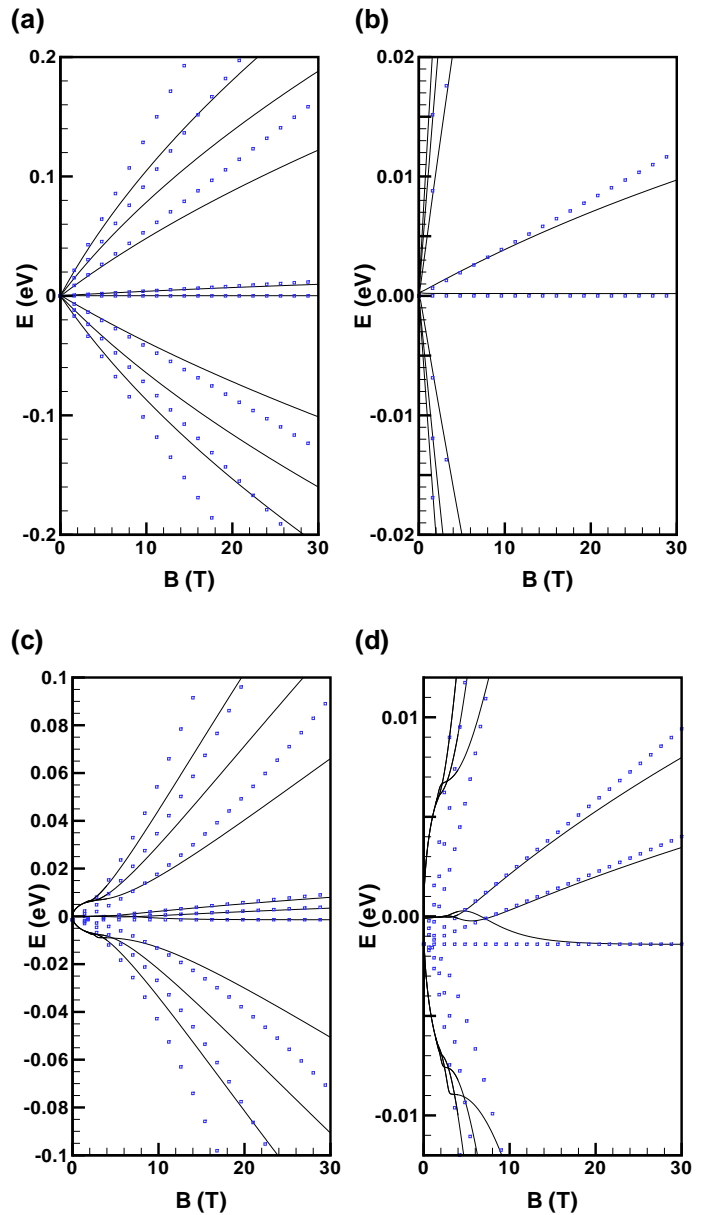}
\caption{(Color online) Behavior of the Landau levels with magnetic field at
zero bias for valley $K_{+}$ for the four-band model (lines) and two-band
model without $\protect\gamma _{2}$ and $\protect\gamma _{3}$ (blue
squares). (a) bilayer graphene; (b) low-energy sector showing the $M=0$
levels of (a); (c) trilayer graphene; (d) low-energy sector showing the $M=0$
levels of (c).}
\label{Fig5}
\end{figure}

In this section, we compute the dispersion of the first few Landau levels
with magnetic field at zero bias. The result is shown in Fig. \ref{Fig5}%
(a),(b) for BLG and Fig. \ref{Fig5}(c),(d) for TLG. The black lines are for
the full (4BM or 6BM) Hamiltonian while the blue squares are the 2BM
dispersions. Figure \ref{Fig5} (b) and (d) shows an enlarged portion of the
low-energy sector that contains the sub-levels in $M=0$ i.e. $n=0,1$ for BLG
and $n=0,1,2$ for TLG.

For BLG, warping term $\gamma _{3}$ in the 4BM\ leads to a very small upward
shift $\sim 10^{-4}$ eV of the $n=0$ level with respect to the 2BM where
this level is exactly at zero energy. (This small shift is also visible in
Fig. \ref{Fig3} (b).) Otherwise, the 2BM\ overestimates the Landau level
energies at all magnetic fields. The energy difference increases with
magnetic field and with Landau level index. For example, at $B=30$ T, the
difference in energy between the 4BM and 2BM is $25\%,35\%$ and $51\%$ for $%
n=1,M=1$ and $M=2$ respectively. For $n=1$, the linear dependence of the
energy predicted by the 2BM is lost at $B\approx 10$ T in the 4BM. For the
higher Landau levels, the accuracy of the 2BM\ is poor excepted at small
magnetic fields.

For TLG, the effects of $\gamma _{2},\gamma _{3}$ are important for $%
B\lesssim 20$ T and, in this range of magnetic field, there is a great
disparity between the full and 2BM even for the $n=0,1,2$ levels in $M=0$
whose energy is otherwise well estimated by the 2BM at higher magnetic
field. The linear dependence of the energy with magnetic field given by the
2BM for $n=0,1,2$ is only present for $B\gtrsim 10$ T in TLG. The energy
difference between the 6BM and 2BM at $B=30$ T is $0.1\%,16\%,18\%,39\%$ and 
$59\%$ for $n=0,1,2$ and $M=1,2$ respectively. As in BLG, the energy
difference between the full and 2BM increases with Landau level index $%
\left\vert M\right\vert $ and with magnetic field.

%%%%%%%%%%%%%%%%%%%%%%%%%%%%%%%%%%%%%%%%%%%%%%%%%%%%%%%%%%%%%%%%%%%%%%%%%
%%%%%%%%%%%%%%%%%%%%%%%%%%%%%%%%%%%%%%%%%%%%%%%%%%%%%%%%%%%%%%%%%%%%%%%%%%

We showed above that, in BLG and TLG, the 2BM is expected to be valid for
magnetic fields $B<56$ T. In the numerical analysis, important quantitative
differences between the 4BM and 2BM appear at magnetic fields much smaller
than this value. In fact, the energy of the Landau level $M=1$ at $B=30$ T
is already near the expected limit of validity of the 2BM which is $E\approx
\gamma _{1}/4$.

For magnetic fields $B>1$ T, the Landau level energies in our calculation
converge for $N_{\max }=100.$ When the magnetic field is decreased, $N_{\max
}$ must be increased, however. In Fig. \ref{Fig5}, we have set $N_{\max
}=900.$ The $\sqrt{B}$ behavior of the Landau level at small magnetic field
in TLG is clearly visible in Fig. \ref{Fig5}(c) and (d). As expected, the
levels with small energy converge more rapidly with $N_{\max }$ than those
with high energy.

\subsection{Dispersion with bias $\Delta _{B}$}

\begin{figure}[tbph]
\includegraphics[scale=1.0]{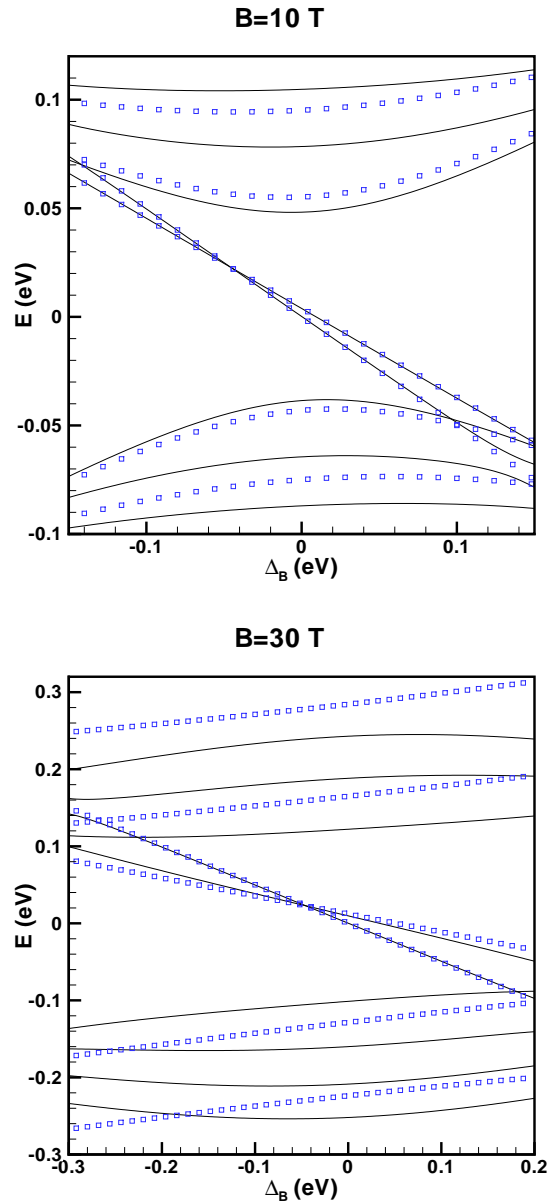}
\caption{(Color online) Dispersion of the first few Landau levels with bias
for the four-band model (lines) and two-band model without $\protect\gamma %
_{3}$ (blue squares) at magnetic field $B=10$ T and $B=30$ T for bilayer
graphene.}
\label{Fig6}
\end{figure}

Figure \ref{Fig6} shows the dependence of the first Landau levels with bias
for BLG for the 4BM (black lines) and the 2BM\ (blue squares) at magnetic
fields $B=10$ T and $B=30$ T. The comparison between the two models is quite
good in BLG for $M=0$ (and even for $M=\pm 1$ at $B=10$ T). For $\left\vert
M\right\vert >1,$ however, the 2BM is quantitatively wrong and does not even
give the correct qualitative dispersion with bias. This is not surprising
since the Landau level energy for $\left\vert M\right\vert =2$ is above $%
\gamma _{1}/4.$

To estimate the range of validity of the 2BM for the $M=0$ levels, we
consider the range of bias where the energies in the two models are close
and where the $M=0$ levels have lower energy than the $M=\pm 1$ levels. For $%
B=10$ T and BLG, this range is roughly given by $\Delta _{B}\in \left[
-0.1,0.1\right] $ eV while for $B=30$ T, it is extended to $\Delta _{B}\in %
\left[ -0.22,0.18\right] .$ Note that a bias $\Delta _{B}=0.1$ eV
corresponds to an electric field of $E=150$ mV/nm which is in the
experimentally accessible range.\cite{Weitz2010}

Because $\gamma _{4}\neq 0$ in our calculation, the two levels in $M=0$ in
BLG meet at a finite negative (positive) bias $\Delta _{B}^{\ast }$ in
valley $K_{+}$($K_{-}$) instead of at $\Delta _{B}^{\ast }=0$. For $\Delta
_{B}<\Delta _{B}^{\ast }$ (valley $K_{+}$) in Fig. \ref{Fig6}, the order of
the $n=0,1$ levels is reversed i.e. $E_{n=1}<E_{n=0}$. From Eqs. (\ref{uno},%
\ref{duo}), $\Delta _{B}^{\ast }=-\delta _{0}-2\gamma _{1}\gamma _{4}/\gamma
_{0}=-0.046$ eV for the values of the hopping parameters considered in this
paper. This change in the relative position of the $n=0,1$ energy levels
leads to interesting physics when Coulomb interaction is introduced because
the Coulomb exchange energy is more negative in level $n=0$ than in level $%
n=1.$ The kinetic energy can thus compete with the exchange energy to create
broken-symmetry states with orbital coherence when $E_{n=1}\leq E_{n=0}.$%
\cite{Lambert2013} This state is characterized by a finite density of
electric dipoles all pointing in the same direction in space.\cite%
{Shizuya2009}

%%%%%%%%%%%%%%%%%%%%%%%%%%%%%%%%%%%%%%%%%%%%%%%%%%%%%%%%%%%%%%%%%%%%%%%%%
%%%%%%%%%%%%%%%%%%%%%%%%%%%%%%%%%%%%%%%%%%%%%%%%%%%%%%%%%%%%%%%%%%%%%%%%%%

\begin{figure}[tbph]
\includegraphics[scale=1.0]{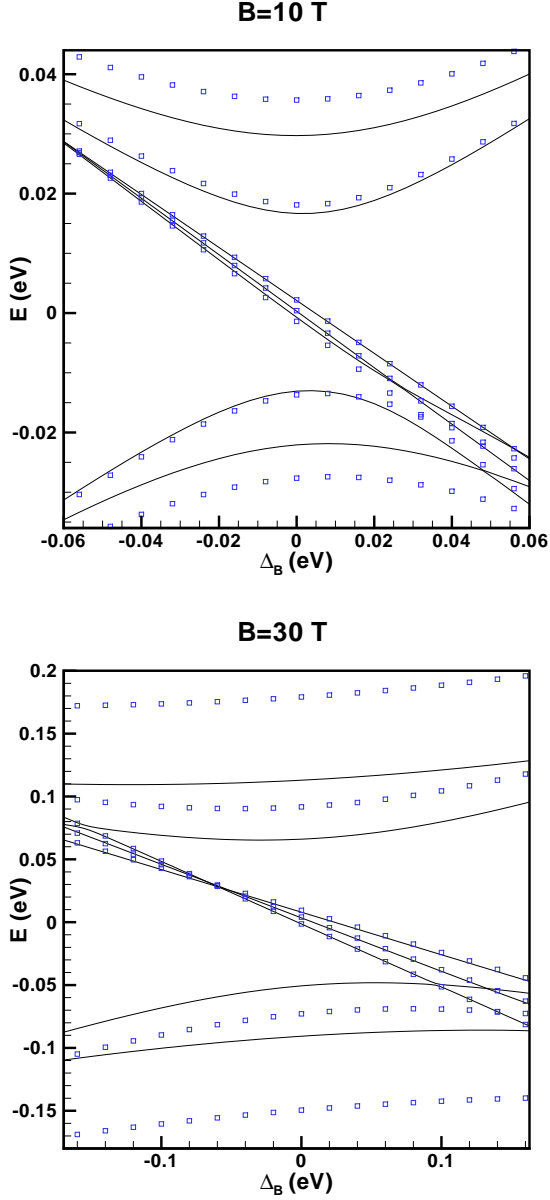}
\caption{(Color online) Dispersion of the first few Landau levels with bias
for the six-band model (lines) and two-band model without $\protect\gamma %
_{2},\protect\gamma _{3}$ (blue squares) at magnetic field $B=10$ T and $%
B=30 $ T for trilayer graphene. }
\label{Fig7}
\end{figure}

\begin{figure}[tbph]
\includegraphics[scale=1.0]{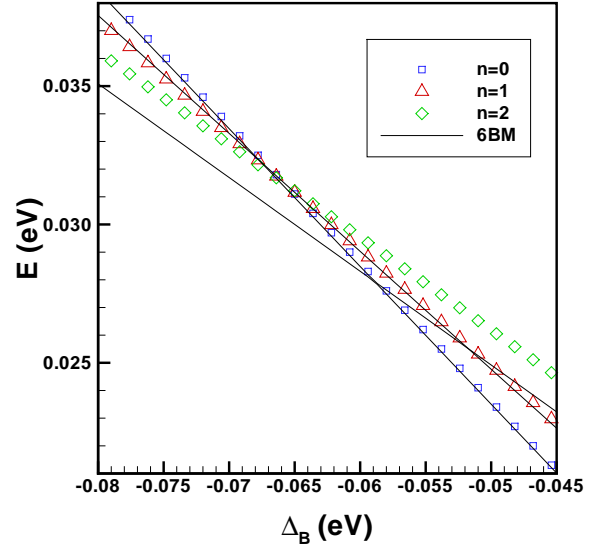}
\caption{(Color online) Dispersion of the three Landau levels $n=0,1,2$ in $%
M=0$ with bias in trilayer graphene at $B=30$ T showing the multiple
crossings in the 6BM (black lines). The symbols are for the 2BM without $%
\protect\gamma _{2},\protect\gamma _{3}.$ }
\label{Fig8}
\end{figure}

Figure \ref{Fig7} shows the dependence of the first Landau levels with bias
in TLG for the 6BM (black lines) and the 2BM\ (blue squares) at magnetic
fields $B=10$ T and $B=30$ T. The $\gamma _{2}$ and $\gamma _{3}$ terms are
important at $B=10$ T and the range of validity of the 2BM\ for the $M=0$
Landau levels is dramatically reduced to $\Delta _{B}\in \left[ -0.032,0.005%
\right] $ eV for $B=10$ T. For $B=30$ T, this range is extended to $\Delta
_{B}\in \left[ -0.14,0.10\right] $ eV. In TLG, as in BLG, the range of
validity increases with magnetic field. In TLG, however, there is a region
around the crossing point of the $n=0,1,2$ levels predicted in valley $K_{+}$
by the 2BM i.e. $\Delta _{B}^{\ast }=2\left( \delta _{0}+2\gamma _{1}\gamma
_{4}/\gamma _{0}\right) =-0.06$ eV, (i.e. in the range $\Delta _{B}\approx %
\left[ -0.07,-0.05\right] $) where the 6BM gives three crossings of the $%
n=0,1,2$ levels (see Fig. \ref{Fig8}). The energy ordering of the levels
with increasing bias is $n=\left( 2,1,0\right) \rightarrow \left(
2,0,1\right) \rightarrow \left( 0,2,1\right) \rightarrow \left( 0,1,2\right)
.$ The multiple crossings disappear if $\gamma _{4}$ is set to zero in which
case there is a single crossing at the value predicted by the 2BM. This
multiple crossing behavior is not captured by the 2BM. When Coulomb
interaction is added to the picture, these multiple crossings should lead to
a variety of broken-symmetry states with orbital coherence and so to a rich
phase diagram for the C2DEG in this range of bias.

%%%%%%%%%%%%%%%%%%%%%%%%%%%%%%%%%%%%%%%%%%%%%%%%%%%%%%%%%%%%%%%%%%%%%%%%%
%%%%%%%%%%%%%%%%%%%%%%%%%%%%%%%%%%%%%%%%%%%%%%%%%%%%%%%%%%%%%%%%%%%%%%%%%%

%%%%%%%%%%%%%%%%%%%%%%%%%%%%%%%%%%%%%%%%%%%%%%%%%%%%%%%%%%%%%%%%%%%%%%%%%
%%%%%%%%%%%%%%%%%%%%%%%%%%%%%%%%%%%%%%%%%%%%%%%%%%%%%%%%%%%%%%%%%%%%%%%%%%

\subsection{Effect of the hopping terms on the Landau level spectrum}

\begin{figure}[tbph]
\includegraphics[scale=1.0]{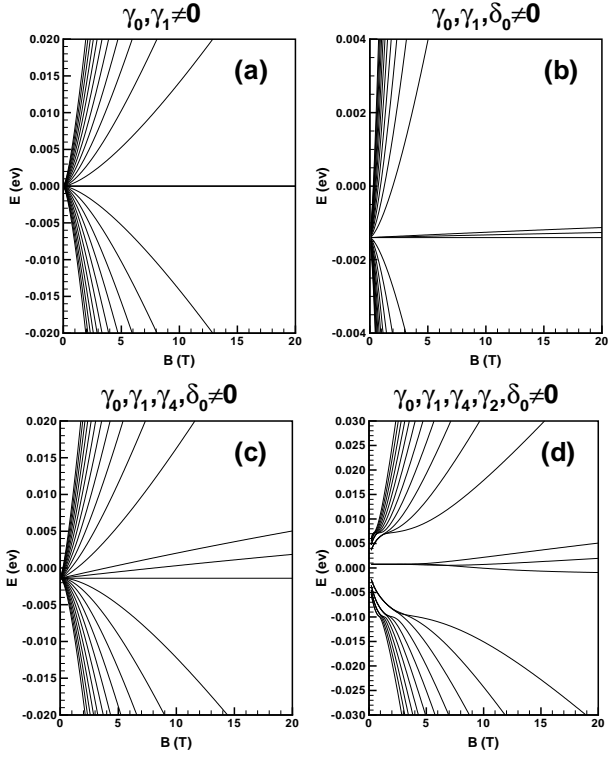}
\caption{Dispersion of the Landau levels with magnetic field at zero bias in
the six-band model of trilayer graphene: (a) minimal model; (b) with the
addition of $\protect\delta _{0};$(c) with the further addition of $\protect%
\delta _{0}$ and $\protect\gamma _{4}$ and (d) with the further addition of $%
\protect\delta _{0},\protect\gamma _{4}$ and $\protect\gamma _{2}.$ }
\label{Fig9}
\end{figure}

It is interesting to see the change in the Landau level spectrum as the
different hopping terms are successively turned on at zero bias in TLG. This
is shown in Fig. \ref{Fig9}. At the top of each figure, we indicate what
hopping terms are non zero. In the minimal model with $\gamma _{0},\gamma
_{1}\neq 0$ only showing that levels $n=0,1,2$ in $M=0$ are degenerate. With 
$\delta _{0}\neq 0$ the spectrum is very slightly shifted downwards since $%
\delta _{0}<0$ and the degeneracy of the $n=0,1,2$ levels is lifted. The
energy of levels $n=1,2$ now depends linearly on the magnetic field (see
Eqs. (\ref{e1}-\ref{e3})). With $\gamma _{4}\neq 0$ the downward shift is
maintained and the slope of the $n=0,1,2$ levels with magnetic field is
increased. With the addition of $\gamma _{2},$ the spectrum is shifted
upwards and crossings occur between the $n=0,1,2$ levels at small magnetic
field (not visible in this figure but noticeable in Fig. \ref{Fig5} (d)).
The dispersion of the higher energy levels with magnetic field changes from
a $\sqrt{B}$ behavior at small $B$ to a $B^{3/2}$ behavior at larger
magnetic field. This occurs at energies $\delta _{0}-\gamma _{2}/2=0.007$ eV
and $\delta _{0}+\gamma _{2}/2=-0.01$ eV. Groups of three levels become
degenerate at very small magnetic field. (See also Refs. %
\onlinecite{Koshino2009,Morimoto2012} where a similar band spectrum is
presented.) Finally, the addition of $\gamma _{3}$ (see Fig. \ref{Fig5} (d))
leads to a small downward shift of the spectrum in Fig. \ref{Fig9} (d) but
does not introduce other significant qualitative changes.

%%%%%%%%%%%%%%%%%%%%%%%%%%%%%%%%%%%%%%%%%%%%%%%%%%%%%%%%%%%%%%%%%%%%%%%%%
%%%%%%%%%%%%%%%%%%%%%%%%%%%%%%%%%%%%%%%%%%%%%%%%%%%%%%%%%%%%%%%%%%%%%%%%%%

\subsection{Eigenstates in the two- and six-band model}

In BLG, the eigenspinors given by Eqs. (\ref{psibilayer}) are coupled
together by the hopping term $\gamma _{3}.$ In order to diagonalize the full
Hamiltonian $H_{\xi },$ we used in Sec. II the basis $\left\{ \Psi
_{N=0,1}^{\left( 0\right) },\Psi _{N=1,1}^{\left( 0\right) },\Psi
_{N=1,2}^{\left( 0\right) },\Psi _{N=1,3}^{\left( 0\right) },\Psi
_{N>1,k}^{\left( 0\right) },...\right\} $ (with $k=1,...,4$) of the
eigenstates of $H_{\xi }^{\left( 0\right) }$ where the spinors $\Psi
_{N,k}^{\left( 0\right) }$ for a given $N$ are ordered is ascending order of
their energy $E_{N,k}^{\left( 0\right) }.$ The eigenstates $\Phi ^{\left(
j\right) }$ of $H_{\xi }$ can be written as linear combinations of the basis
spinors i.e.

\begin{equation}
\Phi ^{\left( j\right) }=\sum_{J\left( N,k\right) }b_{J\left( N,k\right)
}^{\left( j\right) }\Psi _{N,k}^{\left( 0\right) },
\end{equation}%
where the coefficients $b_{k}^{\left( j\right) }$ are obtained by
numerically diagonalizing the full matrix $H_{\xi }.$ If $n_{vec}$
eigenspinors are kept, then $j,J=1,...,n_{vec}$ with the super-index $%
J\left( N,k\right) $ defined by 
\begin{eqnarray}
J\left( 0,1\right) &=&1, \\
J\left( 1,1\right) &=&2, \\
J\left( 1,2\right) &=&3, \\
J\left( 1,3\right) &=&4, \\
J\left( N>1,k\right) &=&4\left( N-1\right) +k.
\end{eqnarray}

In this section, we are interested in comparing the eigenstates of Landau
level $M=0$ in the 4BM or 6BM and 2BM. We start by considering BLG and later
generalize the approach to TLG.

For BLG, the two eigenstates are the $n=0,1$ states that disperse linearly
with bias in Fig. \ref{Fig6}. In the 4BM, we denote these eigenstates by $%
\Phi ^{\left( j_{0}\right) }$ and $\Phi ^{\left( j_{1}\right) }$. In the
absence of coupling due to the warping term, 
\begin{eqnarray}
\Phi ^{\left( j_{0}\right) } &=&\Psi _{N=0,1}^{\left( 0\right)
}\Longrightarrow b_{1}^{\left( j_{0}\right) }=1, \\
\Phi ^{\left( j_{1}\right) } &=&\Psi _{N=1,2}^{\left( 0\right)
}\Longrightarrow b_{3}^{\left( j_{1}\right) }=1,
\end{eqnarray}%
and so the importance of the coupling can be measured by computing the
probabilities%
\begin{eqnarray}
P_{0} &=&\left\vert b_{1}^{\left( j_{0}\right) }\right\vert ^{2}, \\
P_{1} &=&\left\vert b_{3}^{\left( j_{1}\right) }\right\vert ^{2}.
\end{eqnarray}%
The smaller the value of $P_{0},P_{1}$, the more important is the coupling.

In the 2BM, an electron in the $j_{0}$ or $j_{1}$ Landau level is assumed to
reside mostly on site $A_{2}$ (for valley $K_{+}$) and to be described by
the wave function $h_{0,X}\left( \mathbf{r}\right) $ for $j_{0}$ and $%
h_{1,X}\left( \mathbf{r}\right) $ for $j_{1}.$ According to Eq. (\ref{psi2})
in Appendix A, the occupation of the other sites are of order $2$ in the
small quantities defined in this Appendix. In the 4BM, however, the
eigenstate is given by the spinor of Eq. (\ref{psibilayer}) and for level $%
j_{1},$ there may be some electronic amplitude on two other sites of the
unit cell i.e. $B_{2}$ and $A_{1}.$ To measure how well the 2BM describes
the 4BM\ eigenstate in levels $j_{0},j_{1},$ we define the probabilities%
\begin{eqnarray}
\Lambda _{0} &=&P_{0}\left\vert c_{N=0,1,2}\right\vert ^{2}, \\
\Lambda _{1} &=&P_{1}\left\vert c_{N=1,3,2}\right\vert ^{2}.
\end{eqnarray}%
(Note that $\left\vert c_{N=0,1,2}\right\vert ^{2}=1$ so that $P_{0}=\Lambda
_{0}.$)

\begin{figure}[tbph]
\includegraphics[scale=1.0]{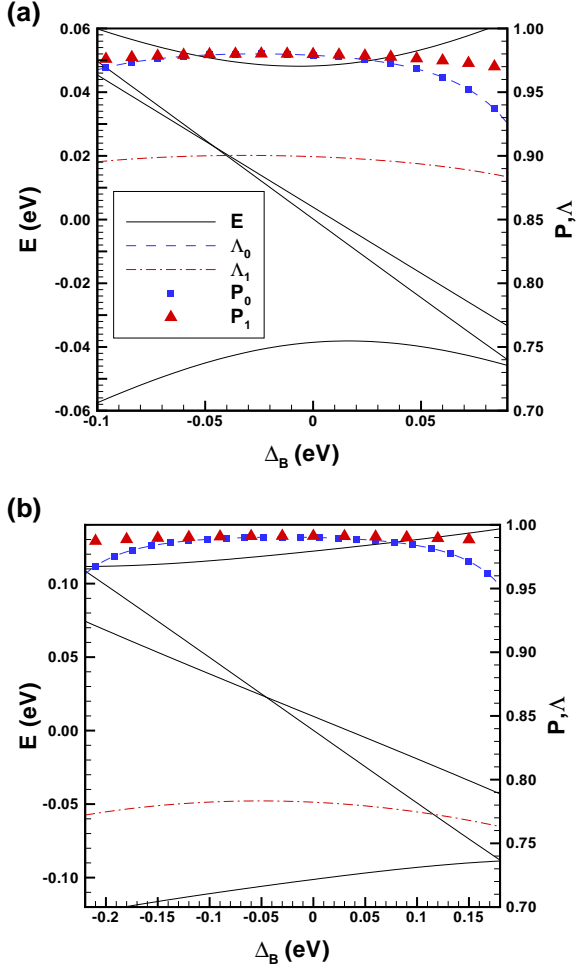}
\caption{(Color online) Low-energy eigenstates of the four-band model and
the probabilities $P$ and $\Lambda $ as a function of bias for the $%
M=0;n=0,1 $ levels of bilayer graphene at (a) $B=10$ T and (b) $B=30$ T.}
\label{Fig10}
\end{figure}

\begin{figure}[tbph]
\includegraphics[scale=1.0]{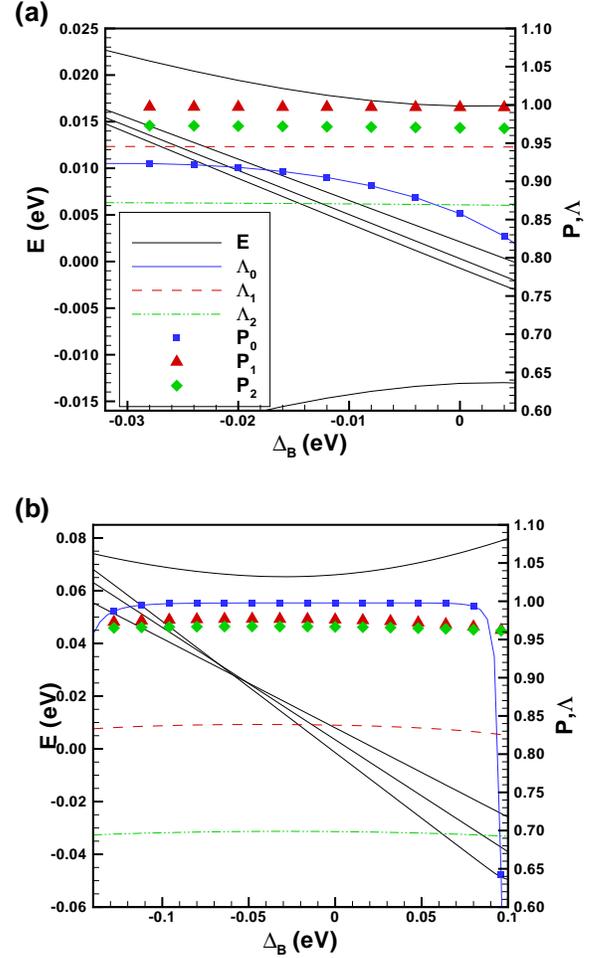}
\caption{(Color online) Low-energy eigenstates of the six-band model and the
probabilities $P$ and $\Lambda $ as a function of bias for the $M=0;n=0,1,2$
levels of trilayer graphene at (a) $B=10$ T and (b) $B=30$ T. }
\label{Fig11}
\end{figure}

We repeat the above procedure for the TLG\ system, defining in a similar way
the probabilities $P_{n}$ and $\Lambda _{n}$ for $n=0,1,2$. Our numerical
results are shown in Fig. \ref{Fig10} for BLG and Fig. \ref{Fig11} for TLG.

Figure \ref{Fig10} shows the probabilities $P_{n}$ and $\Lambda _{n}$ as
well as the dispersion of the $n=0,1$ levels with bias in BLG at magnetic
field of (a) $10$ T and (b) $30$ T. The range of bias is $\Delta _{B}\in %
\left[ -0.1,0.1\right] $ eV for $B=10$ T and $\Delta _{B}\in \left[
-0.22,0.18\right] $ eV for $B=30$ T corresponding to the domain of validity
for the 2BM established in Sec. III (C). In this figure, $P_{0},P_{1}\gtrsim
0.98$ at both magnetic fields and the warping term has very little effect on
the eigenstate. The eigenvectors $\Psi _{N=0,1}^{\left( 0\right) }$ and $%
\Psi _{N=1,2}^{\left( 0\right) }$ are only slightly coupled to the other
eigenstates of $H_{\xi }^{\left( 0\right) }.$ The coupling decreases with
increasing magnetic field as noted before. The probability $\Lambda
_{0}\approx 0.98$ for $B=10$ T and $\Lambda _{0}\approx 0.99$ for $B=30$ T
so that the $n=0$ 2BM eigenvector is a good approximation of the 4BM
eigenvector in this case. For $n=1,\Lambda _{1}\approx 0.90$ for $B=10$ T
and $\Lambda _{1}\approx 0.78$ for $B=30$ T so that the 2BM eigenvector for $%
n=1$ is not as good an approximation as the 2BM eigenvector for $n=0.$

%%%%%%%%%%%%%%%%%%%%%%%%%%%%%%%%%%%%%%%%%%%%%%%%%%%%%%%%%%%%%%%%%%%%%%%%%
%%%%%%%%%%%%%%%%%%%%%%%%%%%%%%%%%%%%%%%%%%%%%%%%%%%%%%%%%%%%%%%%%%%%%%%%%%

Figure \ref{Fig11} shows the probabilities $P_{n}$ and $\Lambda _{n}$ as
well as the dispersion of the $n=0,1,2$ levels with bias in TLG at magnetic
field of (a) $10$ T and (b) $30$ T. The domain of validity of the 2BM is $%
\Delta _{B}\in \left[ -0.032,0.005\right] $ eV for $B=10$ T and $\Delta
_{B}\in \left[ -0.14,0.10\right] $ eV for $B=30.$ In this figure, $%
P_{0},P_{1},P_{2}\gtrsim 0.96$ at $B=30$ T and so both magnetic fields and
the warping term has very little effect in this case. At the lower $B=10$ T
field, however, the coupling is noticeable for level $n=0$. This is what
causes the $n=0$ Landau level to increase in energy for $B\lesssim 10$ T in
Fig. \ref{Fig5} (d). As in BLG, the probabilities $\Lambda _{n}$ decrease
with magnetic field and so the amplitude of the wave function on sites other
than those prescribed by the 2BM are not negligible for $B=30$ T and $n=1,2.$
In the region of the multiple crossings (in the range $\Delta _{B}\in \left[
-0.07,-0.05\right] $), the levels $n=0,1,2$ cross each other but do not
become coupled. In consequence, each level keeps its orbital character in
the sense that the amplitude of the wave function for level $n$ is maximal
in the orbital $h_{n}\left( \mathbf{r}\right) .$

The fact that the electronic amplitude is not small on the high-energy sites
at large magnetic field means that the physical quantities such as electric
current and electromagnetic absorption are not correctly estimated in the
2BM.

%%%%%%%%%%%%%%%%%%%%%%%%%%%%%%%%%%%%%%%%%%%%%%%%%%%%%%%%%%%%%%%%%%%%%%%%%
%%%%%%%%%%%%%%%%%%%%%%%%%%%%%%%%%%%%%%%%%%%%%%%%%%%%%%%%%%%%%%%%%%%%%%%%%%

\section{CONCLUSION}

The two-band model is a useful approximation to study the transport and
optical properties of the chiral two-dimensional electron gas in bilayer and
trilayer graphene but it is important to know its range of applicability. In
this paper, we have compared the Landau level spectrum of the full 6BM (TLG)
and 4BM (BLG) as a function of magnetic field and bias with the predictions
of the 2BM. We can summarize our main conclusions as follows: (a) the 2BM is
generally a better approximation in BLG\ than in TLG; (b) it describes
satisfactorily the $M=0$ Landau levels but differences with the 4BM or 6BM
spectrum increase rapidly with increasing Landau level index $\left\vert
M\right\vert $ and magnetic field; (c) the coupling terms $\gamma _{3}$ in
BLG and $\gamma _{2},\gamma _{3}$ in TLG have little effect on the Landau
level spectrum at magnetic field $B\gtrsim 1$ T in BLG and $B\gtrsim 10$ T
in TLG; (d) at fixed magnetic field and varying bias, the range of validity
of the 2BM increases with magnetic field but is much smaller in TLG\ than in
BLG; (e) for TLG, the multiple crossings between the $M=0$ levels that occur
within a small range of bias in the 6BM are not captured by the 2BM and so
the 2BM may miss some interesting ground states of the C2DEG when Coulomb
interaction is added to the non-interacting Hamiltonian; (f) the amplitude
of the electronic wave function on sites other than those prescribed by the
2BM for the $M=0$ eigenstates increases with magnetic field and is not
negligible at large field; (g) the 2BM consistently overestimates the Landau
level energies.

\begin{acknowledgments}
R. C\^{o}t\'{e} was supported by a grant from the Natural Sciences and
Engineering Research Council of Canada (NSERC). Computer time was provided
by Calcul Qu\'{e}bec and Compute Canada.
\end{acknowledgments}

\appendix

\section{DERIVATION OF THE TWO-BAND MODEL}

Combining the approaches found in Refs. %
\onlinecite{McCannRevue2013,Zhang2010}, we introduce the two-band model in
the following way. The eigenvalue equation to be solved can be written in
the form%
\begin{equation}
\left( 
\begin{array}{cc}
H_{1,1} & H_{1,2} \\ 
H_{2,1} & H_{2,2}%
\end{array}%
\right) \left( 
\begin{array}{c}
\left\vert \Psi _{1}\right\rangle \\ 
\left\vert \Psi _{2}\right\rangle%
\end{array}%
\right) =\varepsilon \left( 
\begin{array}{c}
\left\vert \Psi _{1}\right\rangle \\ 
\left\vert \Psi _{2}\right\rangle%
\end{array}%
\right) ,
\end{equation}%
with the normalization condition%
\begin{equation}
\left\langle \Psi _{1}|\Psi _{1}\right\rangle +\left\langle \Psi _{2}|\Psi
_{2}\right\rangle =1,
\end{equation}%
where $H_{1,1}$ is the Hamiltonian for the low-energy sector, $H_{2,2}$ the
Hamiltonian of the high-energy sector and $H_{1,2},H_{2,1}$ the coupling
terms. In BLG, $H_{i,j}$ are $2\times 2$ matrices and $\Psi _{1},\Psi _{2}$
have $2$ elements. In TLG, $H_{1,1}$ is a $2\times 2$ matrix, $H_{2,2}$ a $%
4\times 4$ matrix and $H_{1,2},H_{2,1}$ are $2\times 4$ and $4\times 2$
matrices respectively and $\Psi _{1}$ has $2$ elements and $\Psi _{2}$ has $%
4.$ We want to derive an equation for the low-energy part $\Psi _{1}$ of the
eigenspinor.

The eigenvalue equation is rewritten as%
\begin{eqnarray}
\left( H_{11}-\varepsilon I\right) \left\vert \Psi _{1}\right\rangle
+H_{12}\left\vert \Psi _{2}\right\rangle &=&0, \\
H_{21}\left\vert \Psi _{1}\right\rangle +\left( H_{22}-\varepsilon I\right)
\left\vert \Psi _{2}\right\rangle &=&0.
\end{eqnarray}%
where $I$ is the unit matrix. Solving for $|\Psi _{2}\rangle $ gives%
\begin{equation}
\left\vert \Psi _{2}\right\rangle =-\left( H_{22}-\varepsilon I\right)
^{-1}H_{21}\left\vert \Psi _{1}\right\rangle
\end{equation}%
so that%
\begin{equation}
\left[ \left( H_{11}-\varepsilon I\right) -H_{12}\left( H_{22}-\varepsilon
I\right) ^{-1}H_{21}\right] |\Psi _{1}\rangle =0.
\end{equation}%
This is an exact equation.

Now, $H_{22}$ can be diagonalized by the transformation 
\begin{equation}
H_{22}=UD_{22}U^{-1}
\end{equation}%
where $D_{22}$ is the diagonal matrix containing the eigenvalues of $H_{22}$%
. It follows that%
\begin{eqnarray}
\left( H_{22}-\varepsilon I\right) ^{-1} &=&\left( U\left(
D_{22}-\varepsilon I\right) U^{-1}\right) ^{-1} \\
&=&U\left( D_{22}-\varepsilon I\right) ^{-1}U^{-1}.  \notag
\end{eqnarray}%
For the low energy excitations, all eigenvalues of $D_{22}>>\varepsilon $ so
that 
\begin{equation}
\frac{1}{d_{22}-\varepsilon }\approx \frac{1}{d_{22}}\left( 1+\frac{%
\varepsilon }{d_{22}}\right) .
\end{equation}%
Hence,%
\begin{equation}
\left( D_{22}-\varepsilon I\right) ^{-1}\approx D_{22}^{-1}\left(
I+\varepsilon D_{22}^{-1}\right) ,
\end{equation}%
and so

\begin{equation}
\left( H_{22}-\varepsilon I\right) ^{-1}\approx H_{22}^{-1}\left(
I+\varepsilon H_{22}^{-1}\right)
\end{equation}%
to first order in $\varepsilon .$

The Schr\"{o}dinger equation now becomes%
\begin{equation}
H_{eff}\left\vert \Psi _{1}\right\rangle =\varepsilon \left\vert \Psi
_{1}\right\rangle ,
\end{equation}%
where%
\begin{equation}
H_{eff}=\left( I+H_{12}H_{22}^{-2}H_{21}\right) ^{-1}\left(
H_{11}-H_{12}H_{22}^{-1}H_{21}\right) .
\end{equation}

We have also,%
\begin{eqnarray}
\left\vert \Psi _{2}\right\rangle &=&-\left( H_{22}-\varepsilon I\right)
^{-1}H_{21}\left\vert \Psi _{1}\right\rangle  \label{psi2} \\
&\approx &-H_{22}^{-1}\left( I+\varepsilon H_{22}^{-1}\right)
H_{21}\left\vert \Psi _{1}\right\rangle .  \notag
\end{eqnarray}

In order to satisfy the normalization condition of the original model, we
define%
\begin{equation}
S=I+H_{12}H_{22}^{-2}H_{21}
\end{equation}%
and make the transformation%
\begin{equation}
\left\vert \widetilde{\Psi }_{1}\right\rangle =S^{1/2}\left\vert \Psi
_{1}\right\rangle .
\end{equation}%
The new eigenvalue equation is 
\begin{equation}
\widetilde{H}_{eff}\left\vert \widetilde{\Psi }_{1}\right\rangle
=\varepsilon \left\vert \widetilde{\Psi }_{1}\right\rangle
\end{equation}%
with%
\begin{eqnarray}
\widetilde{H}_{eff} &=&S^{1/2}H_{eff}S^{-1/2} \\
&=&S^{-1/2}\left( H_{11}-H_{12}H_{22}^{-1}H_{21}\right) S^{-1/2}.  \notag
\end{eqnarray}

Note that using Eq. (\ref{psi2}), we can further approximate%
\begin{equation}
\left\vert \Psi _{2}\right\rangle \approx -H_{22}^{-1}H_{21}\left\vert \Psi
_{1}\right\rangle ,
\end{equation}%
so that it is easy to show that 
\begin{equation}
\left\langle \widetilde{\Psi }_{1}|\widetilde{\Psi }_{1}\right\rangle
\approx \left\langle \Psi _{1}|\Psi _{1}\right\rangle +\left\langle \Psi
_{2}|\Psi _{2}\right\rangle =1.
\end{equation}

The Hamiltonians $H_{eff}$ and $\widetilde{H}_{eff}$ have the same
eigenvalues but $\widetilde{H}_{eff}$ is Hermitian while $H_{eff}$ is not
necessarily Hermitian.

The intralayer hopping $\gamma _{0},\gamma _{1}\,$are much larger than the
small quantities $\left\vert \varepsilon \right\vert ,\left\vert \gamma
_{2}\right\vert ,\left\vert \gamma _{3}\right\vert ,\left\vert \gamma
_{4}\right\vert ,\left\vert \delta _{0}\right\vert ,\left\vert \Delta
_{B}\right\vert $. Keeping only the terms that are linear in the small
quantities and quadratic (BLG) or cubic (TLG)\ in the operators $a,a^{\dag }$
leads to the Hamiltonian for BLG and TLG given in the text. Within this
approximation, $\widetilde{H}_{eff}$ and $H_{eff}$ are identical and the
normalization condition reduces to $\left\langle \Psi _{1}|\Psi
_{1}\right\rangle =1$ meaning that there is no amplitude of the wave
function on the high-energy sites. (From Eq. (\ref{psi2}), the probability
to be on a high-energy site is quadratic in the small quantities.)

\end{document}